\documentstyle[aps]{revtex}
\begin{document}
\draft
\author{B.V.Ivanov}
\title{Expanding, axisymmetric pure radiation gravitational fields with simple 
twist}
\address{Institute for Nuclear Research and Nuclear Energy\\
Tzarigradsko Shausse 72, Sofia 1784, Bulgaria}
\date{24 May 1999}
\maketitle

\begin{abstract}
New expanding, axisymmetric pure radiation solutions are found, exploiting
the analogy with the Euler-Darboux equation for aligned colliding plane
waves.
\end{abstract}


\section{Introduction}

There exist many papers dealing with algebraically special, expanding and
twisting pure radiation solutions of the Einstein equations [1-7]. The
standard form of the metric is [1]: 
\begin{equation}
ds^2=\frac{2d\zeta d\bar \zeta }{\rho \bar \rho P^2}-2\Omega \left(
dr+Wd\zeta +\bar Wd\bar \zeta +H\Omega \right)   \label{1}
\end{equation}
\begin{equation}
\Omega =du+Ld\zeta +\bar Ld\bar \zeta   \label{2}
\end{equation}
Here $r$ is the coordinate along the null congruence of geodesics, $u$ is
the retarded time, the complex coordinates $\zeta ,$ $\bar \zeta $ span a
two-dimensional surface. The metric components are determined by the $r$%
-independent real functions $P,$ $m,$ $M$ and the complex function $L$: 
\begin{equation}
2i\Sigma =P^2\left( \bar \partial L-\partial \bar L\right)   \label{3}
\end{equation}
\begin{equation}
\rho =-\frac 1{r+i\Sigma }  \label{4}
\end{equation}
\begin{equation}
W=\rho ^{-1}L_u+i\partial \Sigma   \label{5}
\end{equation}
\begin{equation}
H=-r\left( \ln P\right) _u-(mr+M\Sigma )\rho \bar \rho +\frac K2  \label{6}
\end{equation}
\begin{equation}
K=2P^2Re\left[ \partial \left( \bar \partial \ln P-\bar L_u\right) \right] 
\label{7}
\end{equation}
where $\partial =\partial _\zeta -L\partial _u$ and $\Sigma $ is the twist.
The basic functions $P,$ $L,$ $m,$ $M$ satisfy the system: 
\begin{equation}
\left( \partial -3L_u\right) \left( m+iM\right) =0  \label{8}
\end{equation}
\begin{equation}
P^{-3}M=Im\partial \partial \bar \partial \bar \partial V  \label{9}
\end{equation}
\begin{equation}
n^2=-2P^3\left[ P^{-3}\left( m+iM\right) \right] _u+2P^3\left( \partial
\partial \bar \partial \bar \partial V\right) _u-2P^2\left( \partial
\partial V\right) _u\left( \bar \partial \bar \partial V\right) _u
\label{10}
\end{equation}
where $V_u=P$,  $n$ is the energy density of pure radiation and the Newton
constant is set to 1. Eqs. (8), (9) and (10) are in fact Eqs. (26.32) and
(26.33) from ref. [1].

It has been noticed in different contexts that the condition $M=0$
(vanishing NUT parameter) simplifies the equations [4,5,6,8,9]. Twisting
gravitational fields with $M=0$ generalize the classes of Robinson-Trautman
[10] and Kerr-Schild fields [11] which are physically realistic, their
simplest representatives being the Schwarzschild, Kerr and Vaidya solutions.

In the present paper we explore this condition applying the method of
Stephani [2]. We discuss axisymmetric fields with the simplest possible
twist. In Sec. II Eqs. (8-10) are reformulated in terms of an invariant
potential which leads to the $L_u=0$ gauge. In Sec. III the main equation
(9) for simplest twist is shown to be equivalent to the Euler-Darboux
equation, which is central in the theory of aligned colliding plane waves
(CPW). We use the known solutions and techniques to find solutions for our
problem. In Sec. IV a closing discussion is presented.

\section{Field equations in the L$_u=0$ gauge}

Following [2] we introduce the invariant complex potential $\phi $ which
solves Eq. (8): 
\begin{equation}
m+iM=\phi _u^3  \label{11}
\end{equation}
\begin{equation}
L=\frac{\phi _\zeta }{\phi _u}  \label{12}
\end{equation}
When $M=0$ we can apply the gauge transformation 
\begin{equation}
u^{\prime }=f\left( u,\sigma \right)   \label{13}
\end{equation}
\begin{equation}
\left( m+iM\right) ^{\prime }=f_u^{-3}\left( m+iM\right)   \label{14}
\end{equation}
to make the mass parameter $m$ a positive or negative constant $m_0$ so that 
\begin{equation}
\phi =m_0^{1/3}\left[ u+iq\left( \sigma \right) \right]   \label{15}
\end{equation}
\begin{equation}
L=i\bar \zeta q\left( \sigma \right) _\sigma   \label{16}
\end{equation}
where $q$ is real and due to the axial symmetry depends only on $\sigma
=\zeta \bar \zeta $. The complex coordinate $\zeta $ is related to the
angular coordinates $\theta ,$ $\varphi $ on the distorted spheres ( $r=r_0,$
$u=u_0$ ) according to 
\begin{equation}
\zeta =\sqrt{2}\tan \left( \theta /2\right) e^{i\varphi }  \label{17}
\end{equation}
Obviously $L_u=0$. This gauge differs from the usual Kerr's gauge [12] $P_u=0
$, but is very suitable when the NUT parameter vanishes. Eqs. (9) and (10)
simplify 
\begin{equation}
\partial \partial \bar \partial \bar \partial V=\bar \partial \bar \partial
\partial \partial V  \label{18}
\end{equation}
\begin{equation}
n^2=6m_0P^{-1}P_u+2P^3\partial \partial \bar \partial \bar \partial
P-2P^2\partial \partial P\bar \partial \bar \partial P  \label{19}
\end{equation}
The second equation is in fact an inequality. When $P_u\neq 0$, $n^2$ can be
made positive by the choice of $m_0$ at least for some region of spacetime
[1,4,8]. The expressions for the metric components simplify too, e.g. the
gauge invariants $\Sigma $ and $K$ read 
\begin{equation}
\Sigma =P^2Q  \label{20}
\end{equation}
\begin{equation}
K=P^2\left( \bar \partial \partial +\partial \bar \partial \right) \ln P
\label{21}
\end{equation}
where $Q=q_{\zeta \bar \zeta }$.

When $m+iM=0$ (Petrov types III and N) Eq. (8) is an identity but still a
potential $\phi $ may be introduced with the property $\partial \phi =0$ and
the subclass of solutions satisfying Eqs. (15) and (16) (with $m_0=1$ ) can
be studied. This results in setting $m_0=0$ in all other equations.

In both cases the main equation (18), which is of fourth order with respect
to $V$, becomes a linear second order equation for $P$. Let us choose the
simplest possible twist $q=\sigma ,$ $Q=1,$ $L=i\bar \zeta $. Then Eqs (18)
and (19) read 
\begin{equation}
\left( \bar \partial \partial +\partial \bar \partial \right) P=0  \label{22}
\end{equation}
\begin{equation}
n^2=6m_0P^{-1}P_u-6P^3P_{uu}-2\sigma ^2P^2\left[ \left( 2P_{uu}+\frac 1\sigma
P_\sigma \right) ^2+4P_{u\sigma }^2\right]   \label{23}
\end{equation}
The last term in Eq. (23) is obviously negative, so necessarily the first
must be positive for a type II solution and the second must be positive for
a type III or N solution.

\section{Reduction to the Euler-Darboux equation}

Let us introduce the complex variable $z=\frac 12\left( \sigma +iu\right) $.
Then Eq. (22) becomes 
\begin{equation}
2\left( z+\bar z\right) P_{z\bar z}+P_z+P_{\bar z}=0  \label{24}
\end{equation}
When $z$ and $\bar z$ are two real variables this is the Euler-Darboux
equation, the main equation in the theory of aligned CPW [13]. We can adapt
the numerous solutions for our problem. We must ensure that $P$ is real and
investigate the regions where $n^2>0$.

In the original variables Eq. (24) may be written as 
\begin{equation}
P_{uu}+P_{\sigma \sigma }+\frac 1\sigma P_\sigma =0  \label{25}
\end{equation}
which is the analog of the equation for vacuum Gowdy cosmologies [14]. Eq.
(25) has been derived from another viewpoint in [8] and discussed there.
Solutions with separated variables behave like $\left( u+c\right) \ln \sigma 
$ or $e^{au}J_0\left( b\sigma \right) $ where $a,$ $b,$ $c$ are constants
and $J_0$ is a Bessel function. The first class includes the simplest CPW
solution $P=-\ln \left( z+\bar z\right) $.

A simple solution is obtained by separating the variables in Eq. (24) [13]: 
\begin{equation}
P=A\left[ \left( a+z\right) \left( a-\bar z\right) \right] ^{-1/2}
\label{26}
\end{equation}
where $A,$ $a$ are constants. $A$ is ignorable, while $a$ must vanish for a
real $P$. Hence 
\begin{equation}
P=B^{-1/2}  \label{27}
\end{equation}
\begin{equation}
B=\sigma ^2+u^2  \label{28}
\end{equation}
The energy density becomes 
\begin{equation}
n^2=-6m_0uB^{-1}-12B^{-3}  \label{29}
\end{equation}
We shall discuss only positive retarded times $u>u_0>0$. The energy density
is regular in $u$ and $\sigma $. If $m_0<0$ and $-m_0>2/u_0^5$ then $n^2>0$.
Type III solutions, however, always have negative energy and are not
realistic. The gauge invariants are also regular and vanish for big $u$%
\begin{equation}
\Sigma =B^{-1}  \label{30}
\end{equation}
\begin{equation}
K=-2\sigma B^{-2}  \label{31}
\end{equation}
The Weyl scalars [1,15] have the following leading terms 
\begin{equation}
\Psi _2=m_0\rho ^3  \label{32}
\end{equation}
\begin{equation}
\Psi _3=-\rho ^2P^3\partial I+O\left( \rho ^3\right)  \label{33}
\end{equation}
\begin{equation}
\Psi _4=\rho P^2I_u+O\left( \rho ^2\right)  \label{34}
\end{equation}
where $I=P^{-1}\bar \partial \bar \partial P$. Plugging Eq. (27) into Eqs.
(33) and (34) we obtain 
\begin{equation}
\Psi _3=-6\rho ^2\zeta \left( \sigma +iu\right) ^2B^{-7/2}+O\left( \rho
^3\right)  \label{35}
\end{equation}
\begin{equation}
\Psi _4=6i\rho \zeta ^2\left( \sigma +iu\right) ^3B^{-4}+O\left( \rho
^2\right)  \label{36}
\end{equation}
It can be shown, using the full expressions for $\Psi _3$ and $\Psi _4$
[15], that the Weyl scalars are regular in $\sigma $ and $u$. When $%
u\rightarrow \infty ,$ $\Psi _3$ and $\Psi _4$ vanish, while $\Psi
_2\rightarrow -m_0/r^3$.

There is an analog of the $\cosh ^{-1}$ solution for CPW. Suppose that $%
P=P\left( y\right) $ where 
\begin{equation}
y=\frac{i\left( \bar z-z\right) }{z+\bar z}=\frac u\sigma   \label{37}
\end{equation}
Then Eq. (24) yields the solution 
\begin{equation}
P=\sinh ^{-1}y  \label{38}
\end{equation}
The energy density is given by the expression 
\begin{equation}
n^2=\frac{6m_0}{uP\left( 1+y^{-2}\right) ^{1/2}}+\frac{6P^3y^3}{u^2\left(
y^2+1\right) ^{3/2}}-\frac{2P^2y^2\left( y^2+4\right) }{u^2\left(
y^2+1\right) }  \label{39}
\end{equation}
The first term is regular in $y$ and positive when $m_0>0$. When $%
y\rightarrow \infty $ the energy density becomes negative and $n\sim -y^2$ $%
\ln ^2y$ no matter how big $m_0$ may be.

Eq. (25) is the starting point for the procedure leading to the first
Yurtsever solution [13,16]. It comprises the expressions 
\begin{equation}
P=B^{l/2}P_l\left( x\right)   \label{40}
\end{equation}
\begin{equation}
P=B^{l/2}Q_l\left( x\right)   \label{41}
\end{equation}
where $P_l$ and $Q_l$ are Legendre functions of the first and second kinds
and $x=uB^{-1/2}$. Solutions with $P_l$ grow infinitely when $\sigma
\rightarrow \infty ,$ $u\rightarrow \infty $ and change sign. A promising
solution is 
\begin{equation}
P=2Q_0\left( x\right) =\ln \frac{1+x}{1-x}  \label{42}
\end{equation}
It's energy density is 
\begin{equation}
n^2=\frac{12m_0}u\left( \frac Px\right) ^{-1}+\frac{12}{u^2}\left( xP\right)
^3-8P^2\frac{4\sigma ^2+u^2}{B\sigma ^2}  \label{43}
\end{equation}
The range of $x$ is $0\leq x<1$ and the first two terms are always positive
when $m_0>0$. The last term has a negative pole for $\sigma \rightarrow 0,$
( $x\rightarrow 1$ ) of the type $\sigma ^{-2}\ln ^2\sigma $ and it can't be
compensated by the positive singularity of the second term which is $\sim
\ln ^3\sigma $. Hence $n^2<0$ for $\sigma \rightarrow 0$.

In [13] a general method is presented for obtaining solutions of the
Euler-Darboux equation. Angular coordinates are introduced, which is
possible for CPW because the variables are bounded. This method does not
have an analog for expanding and twisting solutions.

At last, it should be mentioned that any linear combination of the solutions
derived above is also a solution of Eq. (24).

\section{Discussion}

We have shown that when the NUT parameter vanishes and the gauge $L_u=0$ is
used, the main equation (22) for axisymmetric expanding pure radiation
fields with the simplest twist becomes the second order, linear,
Euler-Darboux equation for $P$. This is the central equation in the theory
of aligned colliding plane waves. We have found analogs of some of the
numerous known solutions adapted to our problem and studied the region of
spacetime where the energy density of pure radiation is positive.

It is interesting that the interaction region of two colliding aligned plane
waves is mathematically equivalent to an expanding distorted spherical wave
with a simple twist. However, this is not the first example of such kind. It
was shown in [17] that the field equations for non-aligned colliding
electro-gravitational plane waves coincide with the Ernst equations [18] for
stationary axisymmetric Einstein-Maxwell fields. The Bell-Szekeres
interacting solution is equivalent to the twistless conformally flat
Bertotti-Robinson solution [13]. Colliding gravitational waves are
algebraically general due to the presence of both $\Psi _1$ and $\Psi _4$.
When an electromagnetic wave collides with a gravitational one sometimes the
solution is of type II [19] without twist and falls in a class of solutions
generalizing the Robinson-Trautman solutions (distorted spherical twistless
waves) [20].

The structure of time-dependent expanding and twisting algebraically special
pure radiation solutions obviously is very rich, because in the case of
simplest possible twist they coincide mathematically with the large variety
of aligned colliding plane waves.

\section*{Acknowledgments}

This work was supported by the Bulgarian National Fund for Scientific
Research under Contract No.F-632.

\end{document}